\documentstyle[twoside,fleqn]{actaps}

\def\day{January 5, 1999}
\def\autor{Boris Tom\'a\v{s}ik, Ulrich Heinz}
\def\shorttitle{Potentials and limits of Bose-Einstein interferometry}

\newcommand{\prc}{Phys.\ Rev.\ C}

\newcommand{\prl}{Phys.\ Rev.\ Lett.\ }
\newcommand{\plb}{Phys.\ Lett.\ B}
\newcommand{\zpc}{Z.\ Phys.\ C}
\newcommand{\epjc}{Eur.\ Phys.\ J.\ C}
\newcommand{\hip}{Heavy Ion Physics}

\newcommand{\sjnp}{Sov.\ J.\ Nucl.\ Phys.\ }

\newcounter{alphnum}

\newenvironment{eqalph}{%
	\setcounter{alphnum}{\value{equation}}%
	\refstepcounter{alphnum}%
	\setcounter{equation}{0}%
	}{%
	\setcounter{equation}{\value{alphnum}}%
	}

\newcommand{\beaa}[1]{\begin{eqalph}\label{#1}}
\newcommand{\eeaa}{\end{eqalph}}
\newcommand{\bea}{\begin{eqnarray}}
\newcommand{\eea}{\end{eqnarray}}

\newcommand{\cy}{{\rm y}}
\newcommand{\cz}{{\rm z}}
\newcommand{\ct}{{\rm t}}
\newcommand{\tcx}{\tilde {\rm x}}
\newcommand{\tcy}{\tilde {\rm y}}
\newcommand{\tcz}{\tilde {\rm z}}
\newcommand{\tct}{\tilde {\rm t}}

\newcommand{\Rs}{R_{s}}		\newcommand{\Rsd}{R_s^2}
\newcommand{\Ro}{R_{o}}		\newcommand{\Rod}{R_{o}^2}
		\newcommand{\Rld}{R_{l}^2}
				\newcommand{\Rold}{R_{ol}^2}
				\newcommand{\Rnd}{R_{0}^2}
	\newcommand{\Rpd}{R_\parallel^2}
\newcommand{\Rt}{R_\perp}	\newcommand{\Rtd}{R_\perp^2}
\newcommand{\vyk}{v_{_{\rm YK}}}\newcommand{\Yyk}{Y_{_{\rm YK}}}
\newcommand{\vykd}{v_{_{\rm YK}}^2}
				\newcommand{\Rmnd}{R_{0}^{\prime 2}}
	
				\newcommand{\Rmpd}{R_\parallel^{\prime 2}}
\newcommand{\Rmt}{R_\perp^\prime}
				\newcommand{\Rmtd}{R_\perp^{\prime 2}}
\newcommand{\vmyk}{v_{_{\rm YK}}^\prime}
				\newcommand{\Ymyk}{Y_{_{\rm YK}}^\prime}
\newcommand{\vmykd}{v_{_{\rm YK}}^{\prime 2}}

\newcommand{\bl}{\beta_l}
\newcommand{\bp}{\beta_\perp}

\newcommand{\la}{\langle}	\newcommand{\lla}{\left \langle}
\newcommand{\ra}{\rangle}	\newcommand{\rra}{\right \rangle}

\renewcommand{\vec}[1]{\mbox{\boldmath$#1$\unboldmath}}

\begin{document}

\title{{POTENTIALS AND LIMITS OF BOSE-EINSTEIN INTERFEROMETRY}}
 
\author{Boris Tom\'a\v{s}ik\email{Boris.Tomasik@physik.uni-regensburg.de}}%
	{Institut f\"ur Theoretische Physik, Universit\"at Regensburg, 
	D-93040 Regensburg, Germany}
\author{Ulrich Heinz\email{Ulrich.Heinz@cern.ch}}{CERN, Theory Division, 
	CH-1211 Geneva 23, Switzerland \\ and
	Institut f\"ur Theoretische Physik, Universit\"at Regensburg, 
	D-93040 Regensburg, Germany}

\abstract{We complete the introduction of Yano-Koonin-Podgoretski\u{\i} (YKP)
parametrization of the correlation function by deriving the corrections
to the correlation radii which arise from releasing the popular smoothness
approximation and approximation of setting the pair momentum on-shell. We 
investigate the definition range of this parametrization and find kinematic
regions in which the YKP parametrization is inapplicable. These problems
disappear if the newly proposed Modified Yano-Koonin-Podgoretski\u{\i}
parametrization is used. We then focus on the physical interpretation
of the correlation radii obtained in the different parametrizations. While
the extraction of the longitudinal source expansion from the YK rapidity 
is found to be rather robust against variations of the source density 
profiles, the extracted emission duration is quite sensitive to such 
variations and cannot be reliably extracted except for the case of extremely
long source lifetimes.}


\begin{flushright}
{\sl
Dedicated to J\'an Pi\v{s}\'ut\\ on occasion of his 60th birthday.}
\end{flushright}

\section{Introduction}
\label{intro}

In recent years Bose-Einstein interferometry \cite{GKW79,BGJ90} for 
ultrarelativistic heavy ion collisions has been developed into a powerful 
analysis tool \cite{U-CRIS,Dronten}. It turned out that not only the sizes 
of boson emitting sources are measurable via the magnitude of the 
so-called {\em correlation radii}, which are the parameters of a Gaussian 
fit to the observed correlation function, but also the {\em source dynamics} 
leaves its fingerprint in their {\em dependence on the average momentum of 
the pair} \cite{MS88,AS96,WSH96,CL96,TH98}. On the other hand it became 
also clear that the temporal and spatial size parameters of the emitter 
are not all separately accessible; only certain combinations of the
space-time variances of the source are measurable \cite{HB95,CSH95a,YKPlett}.
This knowledge is crucial for an understanding of the usefulness of 
various Gaussian parametrizations of the correlator. By an appropriate 
choice of the parametrization one can ensure that the resulting width
parameters reflect specific combinations of the space-time variances
which allow for a particularly straightforward physical interpretation 
in terms of specific source features.

It has recently been argued that for sources with dominant longitudinal
expansion the {\em Yano-Koonin-Podgoretski\u{\i} (YKP) parametrization} 
of the correlator \cite{YK78,Pod83,CNH95} provides a more direct
interpretation of the measurement than the ``traditional'' {\em Cartesian} 
one \cite{Pod83,HB95,CSH95a}. It was stated that the YKP parametrization
allows for direct observation of the longitudinal expansion of the source
and provides a more accurate measurement of the emission duration. These 
arguments were supported by a study involving a model which was believed 
to reproduce the data and to include all relevant features admitted by 
experimental observations \cite{YKPlett,YKPlong}. Of course, strictly 
speaking the ``amount'' of information gained from a measurement is 
independent of the chosen parametrization of the correlator (as long
as it is complete); the extraction of certain quantities of interest
appears, however, easier when using the YKP one. The mathematical
equivalence of the different parametrizations leads, on the other hand,
to relations between their respective correlation radii which must be 
satisfied if the analysis was done correctly. The corresponding relations 
between the Cartesian and YKP radius parameters were given in 
\cite{YKPlett,YKPlong}.

In this paper we complete the program of introducing the YKP parametrization. 
Corrections to the correlation radii due to the off-shellness of the pair 
momentum $K$ and due to the so-called smoothness approximation are given up 
to ${\cal O} (\vec{q}^2)$. (The corresponding corrections for the 
Cartesian parametrization were derived in \cite{CSH95b}.) Moreover, we 
study the definition range of this parametrization. We find that kinematic
regions can exist where it is not defined. A small modification in the 
formulation of YKP parametrization is shown to avoid these problems. 
The cost for this remedy is a slightly less straightforward interpretation
of the resulting radius parameters.

With these improved results at hand we can once more study the question 
of the interpretation of the measured correlation radii in the different
parametrizations. We focus in particular on the extraction of the source 
lifetime (i.e. the emission duration). We find that that the original 
optimistic statements in this context \cite{YKPlett,YKPlong} require
qualification -- the lifetime measurement turns out to be considerably 
more model-dependent than originally anticipated. We illustrate this with 
some simple examples in Section \ref{exam}.

More technical items are deferred into the appendices. In 
Appendix~\ref{discrim} the definition range of YKP parametrization 
is studied. Relations between the correlation radii of different 
parametrizations used in this paper are given in Appendix~\ref{cross}. 
Finally, the mentioned ${\cal O}(\vec q^2)$ corrections to the YKP radius 
parameters are derived in Appendix~\ref{corr}.

\section{Measuring the source}
\label{meas}

Given the emission function $S(x,K)$ (Wigner phase-space density of the 
source) and assuming completely chaotic (independent) production of 
the particles \cite{BDH94}, the correlation function can be calculated via 
\cite{Sur73,Sur74,Pra84,PCZ90,CH94} 
\be
\label{CS}
C(q,K) = 1 + \frac{ | \int d^4x\, e^{i q\cdot x}\, S(x,K) |^2}%
{\int d^4x\, S(x,K + \frac{q}{2}) \int d^4y\, S(y,K - \frac{q}{2})}
\, .
\ee
Here $K$ and $q$ are the average and relative 4-momentum of the particle pair:
\be 
\label{e1}
K = \frac{1}{2} (p_1 + p_2) \, , \qquad    q = p_1 - p_2 \, .  \\
\ee
Equation (\ref{CS}) requires the knowledge of the full quantum mechanical 
structure of the source: the momentum $K$ in the numerator is off-shell. 
In practice, however, when evaluating the correlations it is put to its 
on-shell value by setting
\be
\label{e2}
K^0 \rightarrow E_K = \sqrt{ \vec K^2 + m^2}
\, .
\ee
The formalism also simplifies considerably if both momenta in the 
denominator of (\ref{CS}) are identified with $K$ (the so-called smoothness
approximation). The applied simplifications are expected not to be too 
serious because the interesting region of correlations lies at small 
values of $q$ where the above approximations are good. Nevertheless, 
corrections due to the above simplifications can be (and will be) 
derived, which support these superficial statements in a more rigorous way.

To summarize, the above two approximations lead to the relation
\be
\label{CSos}
C(q,K) \approx 1 + \frac{ | \int d^4x\, e^{i q\cdot x}\, S(x,K) |^2}
{ | \int d^4x\, S(x,K) |^2}
\, ,
\ee
with $K$ taken on-shell according to (\ref{e2}).

It is convenient to parameterize the correlation function in Gaussian form:
\be
\label{e3}
C(q,K) - 1 = \exp ( - q_\mu q_\nu B^{\mu\nu} (K) )
\, .
\ee
Expressions for the elements of the (symmetric) matrix $B$ are obtained by
expanding both (\ref{CSos}) and (\ref{e3}) up to second order in $q$ and 
identifying the coefficients of both series. This leads to 
\be
\label{e4}
B_{\mu \nu} = \la \tilde x_\mu \tilde x_\nu \ra
\, ,
\ee
where
\be
\label{e5}
\tilde x_\mu = x_\mu - \la x_\mu \ra
\, ,
\ee
and $\la \ldots \ra$ denotes the average weighted by the emission function:
\be
\label{e6}
\la f(x) \ra(K) = \frac{\int d^4x\, f(x)\, S(x,K)}{\int d^4x\, S(x,K)} \, .
\ee

One must, however, keep in mind that correlations of two {\em physical}
particles are measured and therefore $q$ and $K$ fulfill the on-shell
constraint
\be
\label{on-shell}
q \cdot K = 0
\, ,
\ee
which reduces the number of independent $q$-components to three. Due to this
only certain combinations of the $B_{\mu \nu}$'s are experimentally 
accessible.

In what follows we focus entirely on a class of azimuthally symmetric
sources representing the fireballs generated in central collisions. It is
common to orient the coordinate system in such a way that the z-axis
(longitudinal, $l$) lies in beam direction and the x-axis (outward, $o$)
in the direction of the transverse component of the pair momentum.
The remaining y-direction is then denoted as sideward ($s$). For 
azimuthally symmetric sources there exists then a symmetry under the 
transformation $\cy \rightarrow -\cy$, leading to only seven non-zero 
components of $B$ of which only four combinations are measurable
\cite{CNH95}. (Note that no such symmetry is present in the outward 
direction as long as the directions parallel and antiparallel to 
$K_\perp$ are not equivalent.) 

In this coordinate frame $q_s$ decouples from the equation (\ref{on-shell}).
When employing this relation in different ways one arrives at different 
Gaussian parametrizations of the correlator.

\subsection{Cartesian parametrization}
\label{cartes}

Resolving the on-shell constraint (\ref{on-shell}) as
\be
\label{e7}
q_t = \bp q_o + \bl q_l
\, ,
\ee
where $\beta_i = K^i / E_K$ and $q_t$ is the zeroth component of relative
momentum, and inserting (\ref{e7}) in (\ref{e3}) leads to the Cartesian 
parametrization of the correlator \cite{Pod83,HB95,CSH95a}
\be
\label{cart-par}
C(q,K) - 1 = \exp [- q^2_s \Rsd (K) - q^2_o \Rod (K) - q^2_l \Rld (K)
- 2 q_o q_l \Rold (K) ]
\, .
\ee
For the correlation radii one obtains
\beaa{e8} \bea
\Rod & = & \la (\tcx - \bp \tct)^2 \ra \, ,  \label{e8a} \\
\Rsd & = & \la \tcy^2 \ra \, ,  \label{e8b} \\
\Rld & = & \la (\tcz - \bl \tct)^2 \ra \, ,  \label{e8c} \\
\Rold & = & \la (\tcx - \bp \tct) ( \tcz - \bl \tct) \ra \, .  \label{e8d} 
\eea \eeaa
Due to the on-shell constraint (\ref{e7}) spatial and temporal source 
sizes are mixed in all observable radius parameters except for $\Rsd$. 
This makes the interpretation of a measurement rather involved since the 
spatial and temporal contributions cannot be resolved in a model-independent 
manner. The situation simplifies for the longitudinal radius $\Rld$ if 
the frame in which analysis is performed is boosted longitudinally such 
that $K_l = \bl = 0$ (Longitudinally Co-Moving System, LCMS) \cite{CP91}. 
Then
\beaa{e9} \setcounter{equation}{2}  \bea
\Rld & = & \la \tcz^2 \ra  \, ,   \label{e9c} \\
\Rold & = & \la (\tcx - \bp \tct)\, \tcz \ra  \, . \label{e9d} 
\eea \eeaa
So the longitudinal dimension, in the longitudinal rest frame of the 
pion pair, is measured directly. The cross-term $\Rold$ may obviously 
take both positive and negative values. The superscript should not confuse 
-- it just expresses the bilinear character (and the dimensionality) of this
parameter.

Access to the lifetime is provided by the difference \cite{CP91}
\be
\label{e10}
R_{\rm diff}^2 = \Rod - \Rsd = \bp^2 \la \tct^2 \ra - 
2 \bp \la \tcx \tct \ra + \la \tcx^2 - \tcy^2 \ra 
\, .
\ee
Of course, one has to assume that $R_{\rm diff}^2$ is dominated by the first
term on the r.h.s., at least at higher $K_\perp$ where the suppression by 
$\bp$ is small. In Section \ref{exam} we investigate this assumption in more
detail.

Note finally that corrections to the correlation radii given by (\ref{e8})
due to the approximation (\ref{CSos}) were derived in \cite{CSH95b} -- 
they are listed in Appendix~\ref{corr}.

\subsection{Yano-Koonin-Podgoretski\u{\i} parametrization}
\label{YKP}

When one wants to measure the source using Bose-Einstein interferometry
the problem of the ``right'' frame for the measurement arises. The 
appropriate choice of frame may even depend on the pair momentum. 
Only a part of the whole (expanding) fireball size is reflected
in the correlation radii, namely that part contributing to production of 
particles of a given momentum \cite{MS88}. In the following this particular 
part of the source will be called {\em effective source}.

Since the measurements are done in order to obtain information on the 
local phase-space characteristics (density, energy density, temperature
etc.) of the emitter one should expect that the appropriate frame for 
the measurement of the (effective) source sizes is its own rest frame. 
Unfortunately, this frame cannot be directly determined in the experiment. 
However, in \cite{Pod83} Podgoretski\u{\i} showed for non-expanding 
sources that the cross-term of the correlator parametrization is a direct
consequence of measuring out of the source rest frame. The issue becomes 
a bit more involved when expanding sources are treated, because then
expansion is present also in the effective source. However, since the 
velocity difference between its edges is not very big, one can still
(at least intuitively) introduce something like an average or effective 
source velocity and try to express the cross-term with its help.

This can be to a large part achieved by resolving the on-shell 
constraint (\ref{on-shell}) as
\be
\label{e11}
q_o = \frac{1}{\bp} q_t - \frac{\bl}{\bp} q_l
\, .
\ee
and rewriting (\ref{e3}) in the so-called {\em Yano-Koonin-Podgoretski\u{\i} 
(YKP) form} \cite{CNH95,YKPlett}:
\be
\!\!\!\!\!\!\!\!\!\!
C(q,K) - 1 = \exp [ - q_\perp^2 \Rtd(K) - (q_l^2{-}q_t^2) \Rpd (K) 
- (\Rnd(K){+}\Rpd (K)) (q \cdot U (K) )^2 ] .
\label{eYKP}
\ee
Here the three correlation radii are manifestly invariant under 
longitudinal boosts, and the cross-term has been absorbed into a 
term involving the Yano-Koonin (YK) velocity $\vyk$ which is the 
longitudinal component of the 4-velocity $U$:
\be
\label{e12}
U = \gamma (1,\, 0,\, 0,\, \vyk) \, , \qquad  
\gamma = \frac{1}{\sqrt{1 - \vykd}}
\, .  \\
\ee
Note also that in the Yano-Koonin-Podgoretski\u{\i} parametrization
the components of the transverse momentum difference occur only
in the combination
\be
\label{e13}
q_\perp = \sqrt{ q_s^2 + q_o^2}
\, .
\ee

For the YKP correlation radii and the YK velocity one can infer 
expressions similar to (\ref{e8}) \cite{YKPlett,YKPlong}:
\beaa{e14} \bea
\Rnd & = & A - \vyk C \, ,  \label{e14a} \\
\Rtd & = & \la \tcy^2 \ra = \Rsd \, ,  \label{e14b} \\
\Rpd & = & B - \vyk C \, ,  \label{e14c} \\
\vyk & = & \frac{A + B}{2C} \left ( 1 - 
\sqrt{ 1 - \left (\frac{2C}{A + B} \right )^2} \right ) \, . \label{e14d} 
\eea \eeaa
Here we introduced the shorthands
\beaa{e15} \bea
A & = & \la \tct^2 \ra - \frac{2}{\bp} \la \tcx \tct \ra + \frac{1}{\bp^2}
\la \tcx^2 - \tcy^2 \ra \, , \label{e15a} \\
B & = & \la \tcz^2 \ra - 2 \frac{\bl}{\bp} \la \tcx \tcz \ra + 
\frac{\bl^2}{\bp^2} \la \tcx^2 - \tcy^2 \ra \, , \label{e15b} \\        
C & = & \la \tct \tcz \ra - \frac{1}{\bp} \la \tcx \tcz \ra - 
\frac{\bl}{\bp} \la \tcx \tct \ra + \frac{\bl}{\bp^2} \la \tcx^2 - \tcy^2 \ra 
\, .  \label{e15c} 
\eea \eeaa
Note that these shorthands would appear as parameters of a (pseudo-) Cartesian
pa\-ra\-me\-tri\-za\-tion with the same $q$-components as used in YKP 
\cite{TH98}. 

Although the YKP radii are invariant under longitudinal boosts, their 
interpretation is connected with the frame where $\vyk = 0$, the so-called
Yano-Koonin frame. In this frame $C$ vanishes. From (\ref{e14a}) and 
(\ref{e15a}) one concludes that $\Rnd$ (and only this parameter) is 
sensitive to the lifetime $\la \tct^2 \ra$ of the source in this frame.
This has led to some optimism in Refs.~\cite{YKPlett,YKPlong} because 
measuring the lifetime in this way seems to be much easier than via 
$R_{\rm diff}^2$ (cf.\ eq.\ (\ref{e10})), due to the missing factor
$\beta_\perp^2$ which may be small. (Note that in the YK frame 
$R_{\rm diff}^2 = \beta_\perp^2 A$.) However, as already mentioned 
in the discussion below (\ref{e10}), both extraction procedures
strongly rely on the assumption that the last two terms in (\ref{e15a})
are small. For small $K_\perp$ this is not so clear, due to the small 
factors $\bp$ in the denominator: for example, although at $K_\perp = 0$ 
we generally expect $\la \tcx^2 - \tcy^2 \ra$ to vanish due to azimuthal 
symmetry, if divided by $\bp^2$ the limit for $K_\perp \to 0$ can be a
not-zero quantity \cite{YKPlong}. 
The true behaviour of $A$ for small but nonzero 
$K_\perp$ can thus only be clarified by model studies. For large 
$K_\perp$, on the other hand, different models give different 
predictions for $\la \tcx^2 - \tcy^2 \ra$ (see next Section). 
Unless $\la \tct^2 \ra$ becomes very large (much larger than the 
geometric contributions $\la \tcx^2 \ra$, $\la \tcy^2 \ra$), which may 
happen during compound nucleus formation at low energies \cite{Lisa93}
but is unlikely for the rapidly disintegrating sources formed in 
relativistic heavy ion collisions, the extraction of the lifetime 
thus appears to be quite model-dependent. This reduces the power of 
$\Rnd$ (and $R_{\rm diff}^2$ as well) for measuring the emission duration. 
In fact, $\Rnd$ may even become negative due to the discussed ``correction 
term'' \cite{U-CRIS}.

On first sight, since (\ref{e15b}) has a similar structure as (\ref{e15a}),
one might conclude that the same problems must appear when extracting the 
longitudinal source size from $\Rpd$. Fortunately, here the situation
is better since now the sub-leading terms are multiplied with $\bl$ and 
$\bl^2$, respectively. In the YK frame this velocity is usually small
(see below). Although this is again a model-dependent statement, it 
appears to be valid for all models which are in agreement with the
data. Thus $\Rpd$ is a good estimator for the longitudinal dimension
of the source measured in YK frame.

\subsection{Modified Yano-Koonin-Podgoretski\u{\i} parametrization}
\label{MYKP}

So far, nothing has been said about the definition range and the resulting
applicability of YKP parametrization. Clearly, it is not defined for
$K_\perp = \bp = 0$ as the on-shell constraint (\ref{e11}) is singular
at this point. However, a more serious problem is that the argument of 
the square root in (\ref{e14d}) can become negative 
(see Appendix~\ref{discrim}). Then the YKP parametrization is
{\em ill-defined} (the YKP parameters become complex and thus unphysical). 
We hasten to stress that {\em the applicability region of the YKP 
parametrization cannot be checked directly by the experiment.} It is 
therefore strongly recommended to cross-check the correlation radii 
obtained from a YKP fit via the relations (\ref{YKP2BP})/(\ref{BP2YKP}) 
and/or (\ref{MYKP2YKP})/(\ref{YKP2MYKP}).  

As shown in Appendix \ref{discrim}, the failure of the YKP parametrization 
is due to the use of $q_\perp$ instead of $q_s$ as an independent
variable. In other words, the elimination of $q_o$ via (\ref{e11}) is
not done completely. This leads to the appearance of contributions 
from $\la \tcy^2 \ra / \bp^2$ in (\ref{e15}). One can, however, eliminate 
$q_o$ completely without giving up the longitudinal boost-invariance
of the fitted radius parameters by writing the correlator in the form
\bea
\nonumber
C(q,K) - 1 & = & \exp [ - q_s^2 \Rmtd (K) - (q_l^2 - q_t^2) \Rmpd (K) \\&&
- (\Rmnd (K) + \Rmpd (K) ) (q \cdot U' (K))^2 ]
\, ,
\label{e16}
\eea
with
\be
\label{e17}
U' = \gamma' (1,\, 0,\, 0,\, \vmyk) \, , \qquad \gamma' = 
\frac{1}{\sqrt{1 - \vmykd}}
\, .
\ee
This will be called the {\em Modified Yano-Koonin-Podgoretski\u{\i}} 
(YKP') parametrization. For its correlation radii one finds
\beaa{e18} \bea
\Rmnd & = & A' - \vmyk C' \, ,  \label{e18a} \\
\Rmtd & = & \la \tcy^2 \ra = \Rtd = \Rsd \, ,  \label{e18b} \\
\Rmpd & = & B' - \vmyk C' \, , \label{e18c} \\
\vmyk & = & \frac{A' + B'}{2 C'} 
\left ( 1 - \sqrt{ 1- \left ( \frac{2 C'}{A' + B'} \right )^2 } 
\right ) \, ,   \label{e18d} 
\eea \eeaa
where
\beaa{e19} \bea
A' & = & \la \tct^2 \ra - \frac{2}{\bp} \la \tcx \tct \ra + \frac{1}{\bp^2}
\la \tcx^2 \ra = A + \frac{1}{\bp^2} \la \tcy^2 \ra \, ,  \label{e19a} \\
B'& = & \la \tcz^2 \ra - 2 \frac{\bl}{\bp} \la \tcx \tcz \ra + 
\frac{\bl^2}{\bp^2} \la \tcx^2 \ra = B + \frac{\bl^2}{\bp^2} \la \tcy^2 \ra
\, ,  \label{e19b} \\
C' & = & \la \tct \tcz \ra - \frac{1}{\bp} \la \tcx \tcz \ra - 
\frac{\bl}{\bp} \la \tcx \tct \ra + \frac{\bl}{\bp^2} \la \tcx^2 \ra = 
C + \frac{\bl}{\bp^2} \la \tcy^2 \ra \, .  \label{e19c} 
\eea \eeaa

The interpretation of $\Rmnd$ and $\Rmpd$ is again easiest in the
frame where $\vmyk = 0$, the ``Modified Yano-Koonin (YK') frame''.
It is not as straightforward as in case of the original YKP radii. 
This is the price for a parametrization which is defined everywhere. 
While in YKP parametrization one could hope that the last ``correction 
terms'' of (\ref{e15a}), (\ref{e15b}), and (\ref{e15c}) are small, 
now only the (large) term $\la \tcx^2 \ra$ appears instead of the 
(smaller) difference $\la \tcx^2 - \tcy^2 \ra$. Therefore, the deviations 
of $\Rmnd$ and $\Rmpd$ from $\la \tct^2 \ra$ and $\la \tcz^2 \ra$, 
respectively, are larger than those of $\Rnd$ and $\Rpd$. Still, 
for the measurement of the longitudinal size $\la \tcz^2 \ra$ the 
situation may not be bad because $\la \tcx^2 \ra$ in (\ref{e19b}) is 
again multiplied by $\bl^2$, which is small in the YK' frame, exactly 
like discussed above in case of $\Rpd$. A measurement of 
$\la \tct^2 \ra$ via $\Rmnd$, on the other hand, is definitely impossible.

A good feature which is preserved in the Modified YKP parametrization is 
that the velocity $\vmyk$ still maps to a very good approximation the 
velocity of the effective source. This will be demonstrated in the
next Section.

Finally, we note that while the relative momentum components used in the 
original YKP parametrization satisfy the inequality
\be 
\label{e20}
q_\perp \geq q_o = \frac{1}{\bp} q_t - \frac{\bl}{\bp} q_l
\, , 
\ee
which means that the data points never fill the whole three-dimensional 
$q$-space, no such restriction exists for $q_s$ which is used in the 
modified YKP parametrization. This should help to avoid certain 
technical problems in the fitting procedure which can occur with 
the YKP parametrization.

\paragraph{Cross-relations}
Since all parametrizations discussed so far are just different 
representations of the same correlation function, there must exist 
relations between their correlation radii. They are listed in 
Appendix~\ref{cross}.

These relations provide a powerful check of the correctness of results 
obtained from fitting the measured correlation function with Gaussian 
parametrizations; if Cartesian and YKP (and eventually Modified YKP)
fits are performed independently, the resulting correlation radii must
fulfill the corresponding cross-check relations. Such control is 
{\em strongly recommended in order to avoid the pitfalls related to 
the possible non-existence of the YKP parametrization} for the data 
under study (which may not show up clearly in the fitting process but 
might cause it to converge to a wrong result).  

\paragraph{Off-shell corrections}
Before closing this section we study the corrections to the YKP and 
Modified YKP correlation radii connected with the approximations made 
in (\ref{CSos}). They are easily inferred by combining the known 
expressions (\ref{ctCartd}) and (\ref{ctCartn}) according to (\ref{BP2YKP}) 
and (\ref{BP2MYKP}). The resulting corrections are given in
Appendix~\ref{corr}. In the case of the Modified YKP parametrization
an interesting phenomenon is observed: while the correction terms 
accounting for the release of the smoothness approximation 
$P_1(p_1)P_1(p_2) \approx [P_1(K)]^2$ in the denominator (where 
$P_1(p)= \int d^4x\, S(x,p)$ is the invariant single-particle spectrum) 
and those accounting for the release of the on-shell approximation 
$K^0\approx E_K$ in the numerator diverge separately in the limit 
$K_\perp \to 0$, this singularity cancels in their sum. For the YKP 
parametrization the singularities already cancel at the level of the 
individual corrections. The reason for this difference is the use of 
$q_\perp$ in the YKP parametrization (instead of $q_s$ as in the modified 
one), leading to the appearance of $R_{\rm diff}^2$ in (\ref{BP2YKP}) 
(instead of $\Rod$ in (\ref{BP2MYKP})).

\section{Examples}
\label{exam}

In this Section we illustrate some statements made in the previous Section 
with the help of a model for a longitudinally and transversely expanding 
fireball which is locally thermalized. This is neither an extensive model 
study nor an analysis of real data. Therefore we will stay on a rather 
superficial level and won't go very deeply into the model structure.

Our model (in fact it is a class of models) is expressed by the emission 
function \cite{CL96,TH98,YKPlong}
\bea
S(x,K)\, d^4 x & = & \frac{1}{(2\pi)^3} M_\perp \cosh (Y - \eta)\, 
\exp \left ( - \frac{K \cdot u(x)}{T} \right )\nonumber \\
&& \times  G(r)\, 
\exp \left ( - \frac{(\eta - \eta_0)^2}{2\, \Delta \eta^2} \right ) \nonumber 
\\ && \times
\frac{\tau\, d \tau}{\sqrt{2 \pi\, \Delta \tau^2}} \,
\exp \left ( - \frac{(\tau - \tau_0)^2}{2\, \Delta \tau^2} \right )\, d\eta\,
r\, dr\, d\varphi \, .
\label{e21}
\eea
The momentum is parameterized here in terms of transverse mass 
$M_\perp = \sqrt{K_\perp^2 + m^2}$, rapidity $Y$, and the transverse 
momentum $K_\perp$ such that
\be
\label{e22}
K = (M_\perp \cosh Y,\, K_\perp,\, 0,\, M_\perp \sinh Y)  
\, .
\ee
As space-time coordinates in the transverse plane we use the usual polar 
coordinates $r$ and $\varphi$, and the remaining two directions are 
parameterized by longitudinal proper time $\tau = \sqrt{\ct^2 - \cz^2}$ 
and space-time rapidity $\eta = 0.5 \ln [(\ct + \cz) / (\ct - \cz) ]$. 
The $\eta$-profile is chosen to be Gaussian with width $\Delta\eta$ and 
the center at $\eta_0$. The freeze-out times are distributed by a Gaussian 
of width $\Delta\tau$ centered around the average freeze-out time 
$\tau_0$. 

We will study models with two different transverse profiles: a
{\em Gaussian} one
\be
\label{e23}
G(r) = \exp \left ( - \frac{r^2}{2\, R_G^2} \right )
\, ,
\ee
and a {\em box-shaped} one
\be
\label{e24}
G(r) = \theta (R_B - r)
\, .
\ee

The dynamics of the source is implemented via the velocity field $u(x)$,
\be
\label{e25}
u(x) = ( \cosh \eta_t \, \cosh \eta,\, \cos \varphi \, \sinh \eta_t, \,
\sin \varphi \, \sinh \eta_t, \, \cosh \eta_t \, \sinh \eta)
\, ,
\ee
where the transverse expansion rapidity $\eta_t$ is assumed to scale
linearly with the radial distance $r$:
\be
\label{e26}
\eta_t = \eta_f \frac{r}{r_{\rm rms}}
\, .
\ee
$\eta_f$ is the scaling factor. Note that the transverse rms radius for 
the Gaussian source is
\be
\label{e27}
r_{\rm rms} = \sqrt{2} R_G
\, ,
\ee
while for the box-shaped transverse profile (\ref{e24}) one has
\be
\label{e28}
r_{\rm rms} = \frac{R_B}{\sqrt{2}}
\, .
\ee
The same rms radius is thus ensured by setting $R_B=2R_G$. The source 
dynamics shows up in the coupling of the velocity field (\ref{e25}) with 
the momentum (\ref{e22}) via the Boltzmann factor $\exp [ - K \cdot u /T]$;
$T$ is the freeze-out temperature. Using (\ref{e22}) and (\ref{e25}) the 
scalar product in the exponent is written as
\be
\label{e29}  
K \cdot u = M_\perp\, \cosh(Y - \eta)\, \cosh\eta_t - K_\perp\, \cos\varphi\,
\sinh\eta_t \, .
\ee

For our calculation we take the data-inspired parameter values listed in 
Table~1.
\begin{table}
\begin{center}
\caption{{\bf Table 1 \ \ } 
Values of model parameters used in the calculation}
\begin{tabular}{lcc}
\hline\hline
  & box & Gauss \\
\hline
temperature $T$ &  \multicolumn{2}{c}{0.1 GeV} \\
transverse flow scaling parameter $\eta_f$ & 0.57  & 0.64 \\
geometric transverse radius $R_B$/$R_G$ & 13 fm  &  6.5 fm  \\
average freeze-out proper time $\tau_0$ & \multicolumn{2}{c}{7.5 fm/$c$} \\
mean proper emission duration $\Delta\tau$ & \multicolumn{2}{c}{2 fm/$c$} \\
width of the space-time rapidity profile $\Delta\eta$ &
\multicolumn{2}{c}{1.3} \\ 
\hline
pion mass & \multicolumn{2}{c}{0.139 GeV/$c^2$} \\
kaon mass & \multicolumn{2}{c}{0.493 GeV/$c^2$} \\
\hline \hline
\end{tabular}
\end{center} 
\end{table}
They are not obtained from a careful fit to the data, but seem to reproduce
the measured correlations from Pb+Pb collisions at 160~$A$GeV/$c$ 
\cite{App98,Diss} reasonably well. Note that the different values of 
$\eta_f$ used in the two models lead to the {\em same average transverse 
expansion velocity} $\bar v_\perp$ in both cases. This allows for a 
better comparison of the results, since $\bar v_\perp$ appears to be 
the relevant (model-independent) physical quantity expressing the 
strength of the transverse collective expansion. The transverse sizes 
$R_G$ and $R_B$ are chosen such that the resulting rms radii in both 
models are equal.

In order to provide the reader with a more intuitive picture we show in
Fig.~1 transverse cuts of these sources as seen by midrapidity pions of 
transverse momentum $K_\perp=0.4$ GeV/$c$.
\begin{figure}[t]
\begin{center}\mbox{\input epsf \epsfysize 6.5cm
                        \epsfbox{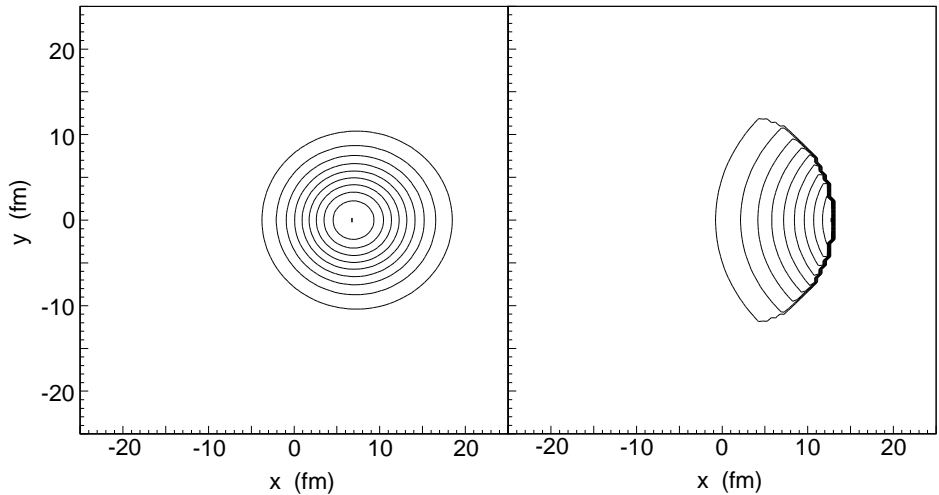}}\end{center}
\caption{{\bf Figure 1\ \ } 
Transverse cuts of the two models from Table 1 at $\eta_0 = 0$
for midrapidity pions ($Y=0$) of transverse momentum $K_\perp = 0.4\, 
\mbox{GeV}/c$. Left: Gaussian source; right: box-shaped source.
(Note that the ``corners'' of the box-shaped source are a plotting
artifact.)}
\end{figure}
The different shape of the effective sources obtained from the two models 
is crucial for an understanding of the $M_\perp$-dependences of the 
correlation radii. These will be studied in what follows. 

\subsection{$M_\perp$-dependence of the correlation radii}
\label{Mperp}

In Figs.~2 and 3 we show 
\begin{figure}[t]
\begin{center}\mbox{\input epsf \epsfysize 10cm
                        \epsfbox{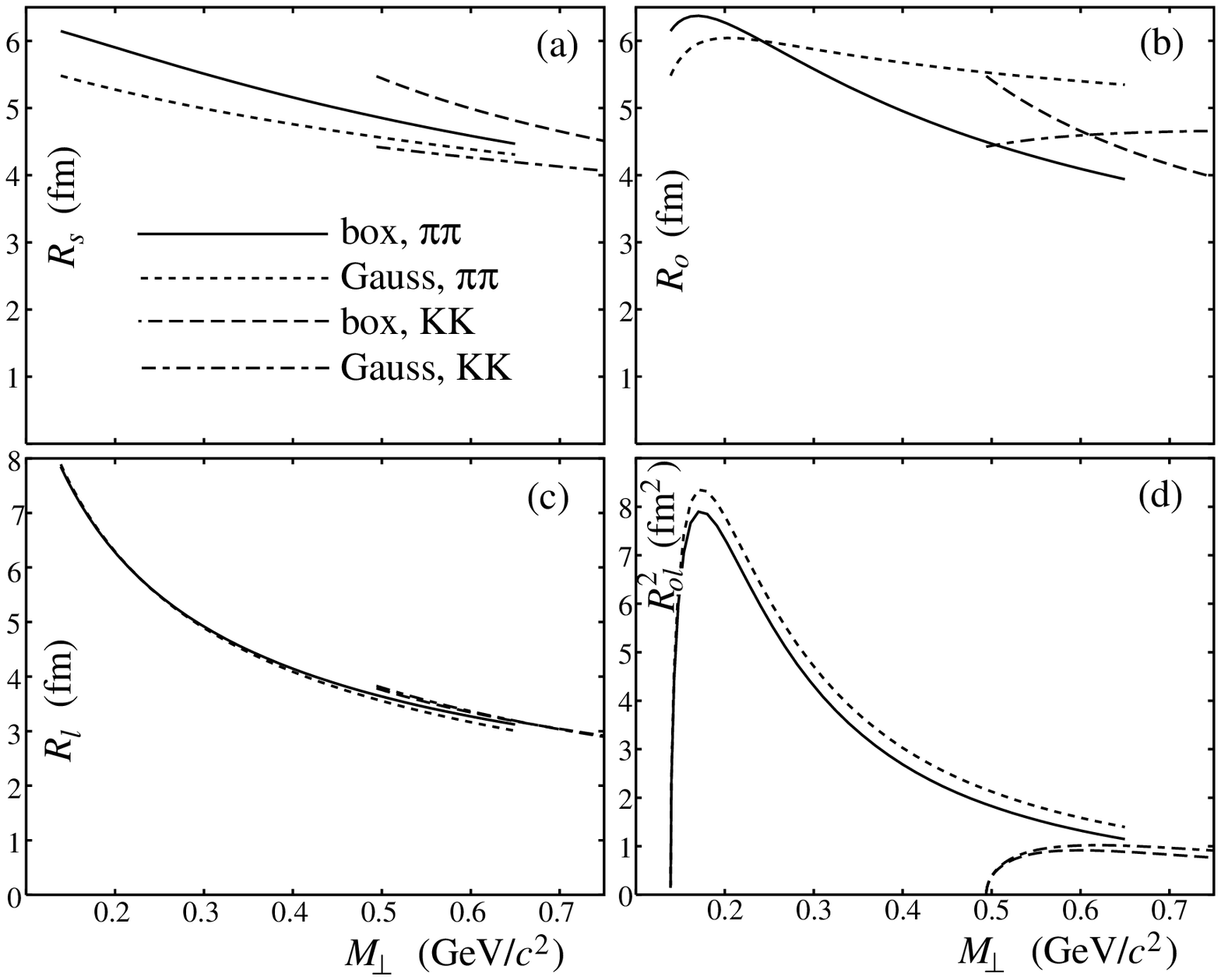}}\end{center}
\caption{{\bf Figure 2\ \ } 
Correlation radii of the Cartesian parametrization for the models 
introduced in (\ref{e21}) with parameter values from Table~1. The 
calculation is performed in the LCMS for particle pairs at slightly forward 
rapidity ($\eta_0 = -1$). Both pion (solid and dotted lines) and kaon
(dashed and dash-doted lines) correlations were computed. Transverse 
profiles: box (solid, dashed), Gauss (dotted, dash-dotted).}
\end{figure}
\begin{figure}[t]
\begin{center}\mbox{\input epsf \epsfysize 6.45cm
                        \epsfbox{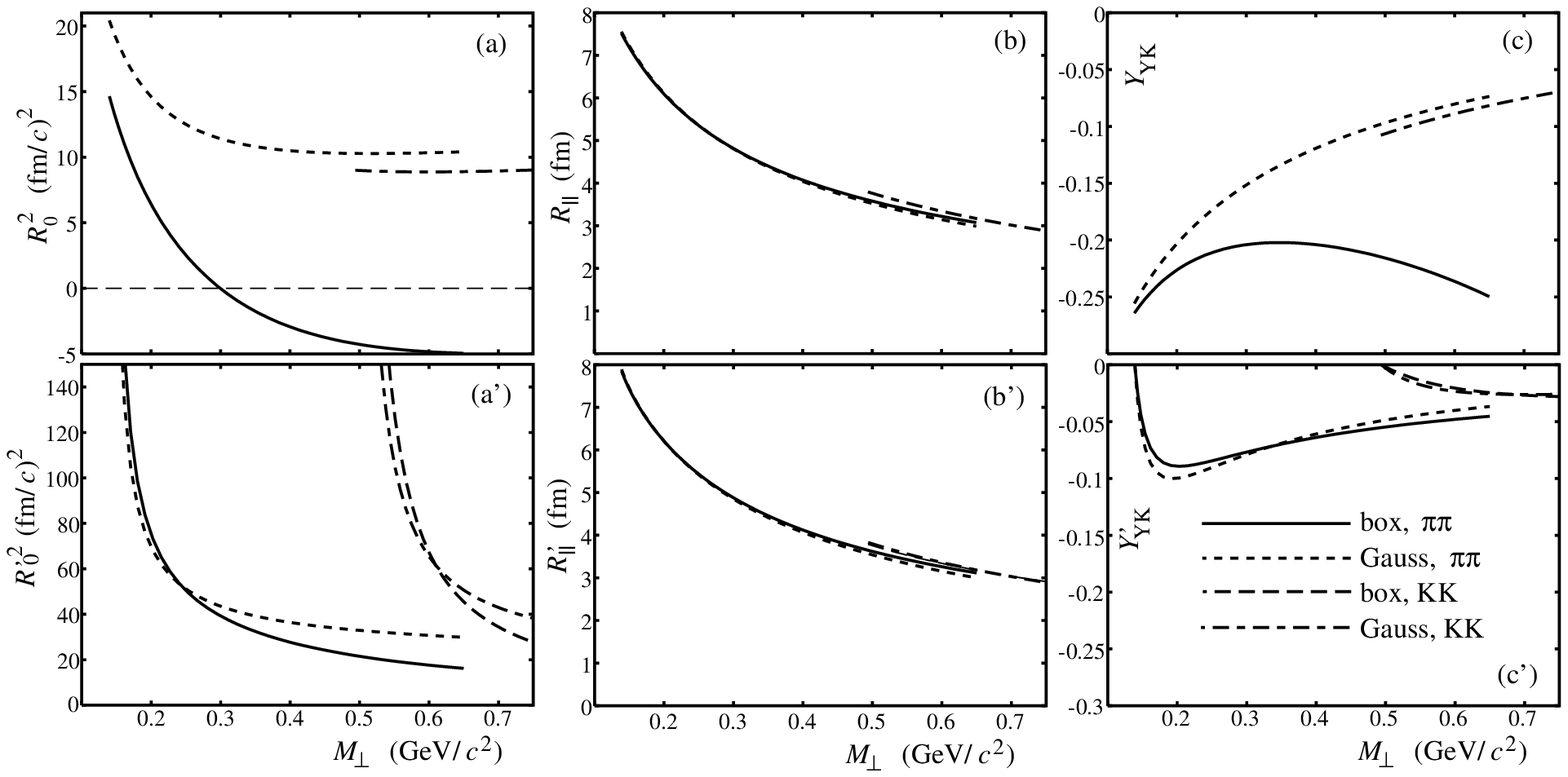}}\end{center}
\caption{{\bf Figure 3\ \ }
YKP (upper row) and Modified YKP (lower row) radii for the same models and 
the same kinematics as in Fig.~2. Note that $\Rt = \Rmt = \Rs$.
Line symbols are as in Fig.~2. At the given kinematics no YKP 
parametrization exists for kaons from the box-shaped source.}
\end{figure}
the $M_\perp$-dependences of the correlation radii of all mentioned 
parametrizations. We evaluate the radii in the LCMS, and in order to get 
a non-vanishing cross-term and Yano-Koonin velocity we take particles at 
slightly forward rapidity $(Y = 0,\, \eta_0 = -1)$. We compute both pion
and kaon correlations. Instead of plotting the Yano-Koonin velocity in 
Fig.~3c we show the Yano-Koonin rapidity
\be
\label{e30}
\Yyk = \frac{1}{2} \ln \frac{1 + \vyk}{1 -\vyk}
\, .
\ee
In Fig.~3c' we show the modified YK rapidity, defined via an analogous 
relation involving $\vmyk$.

Let us comment on a few characteristic features. Although the average 
transverse expansion velocity is the same in both models, the box-shaped 
source generates a stronger $M_\perp$-dependence of $R_s$ than the 
Gaussian one. This effect is even stronger for $\Ro$. Moreover for 
higher transverse mass $\Ro$ is smaller than $\Rs$ for the box-shaped 
source while the opposite is true for the Gaussian source. This results 
from the different shapes of the effective sources shown in Fig.~1: 
while for the Gaussian transverse profile the effective source for higher
$M_\perp$ expands into the dilute tail of the density distribution,  
for the box-profile it is squeezed towards the outer boundary, leading 
to much smaller widths, especially in the outward direction. Note that 
this has important phenomenological consequences: even if the temperature 
is known, the exact amount of transverse flow cannot be inferred uniquely
from the shape of $\Rs(M_\perp)$ without making assumptions about the 
transverse source geometry. By combining measurements of $\Rs$ and 
$\Ro$, however, one should be able to shed some light on this question. 
Definitely, this problem will deserve a more detailed study when fitting 
real data.

The mentioned behaviour is also reflected in the $M_\perp$-dependence 
of $\Rnd$. For a box-shaped source $\la \tcx^2 \ra < \la \tcy^2 \ra$;
via the last term in (\ref{e15a}) this leads to negative values of
$\Rnd$. For kaons from this source the YKP parametrization does not even
exist in the studied kinematic region.

This last problem does not exist in the Modified YKP parametrization, as 
seen in Fig.~3. The behaviour of $\Rmnd$ is completely dominated by the
last term in (\ref{e19a}). The comparison of both models is in accord
with Fig.~2.

Since the longitudinal geometry and dynamics are identical in both models, 
no differences are observed in the behaviour of the longitudinal correlation 
radii. Even in case of $\Rmpd$ the correction terms (the last two terms 
in (\ref{e19b})) lead to negligible effects, since they are multiplied 
by $\bl^2$ which is small in the YK frame.

\subsection{YK rapidity and source velocity}
\label{YKrap}

Since the direct access to the effective source velocity is probably the 
main strength of the YKP/Modified YKP parametrization, Fig.\ 4 is 
entirely devoted to the investigation of this particular feature.
\begin{figure}[t]
\begin{center}\mbox{\input epsf \epsfysize 10cm
                        \epsfbox{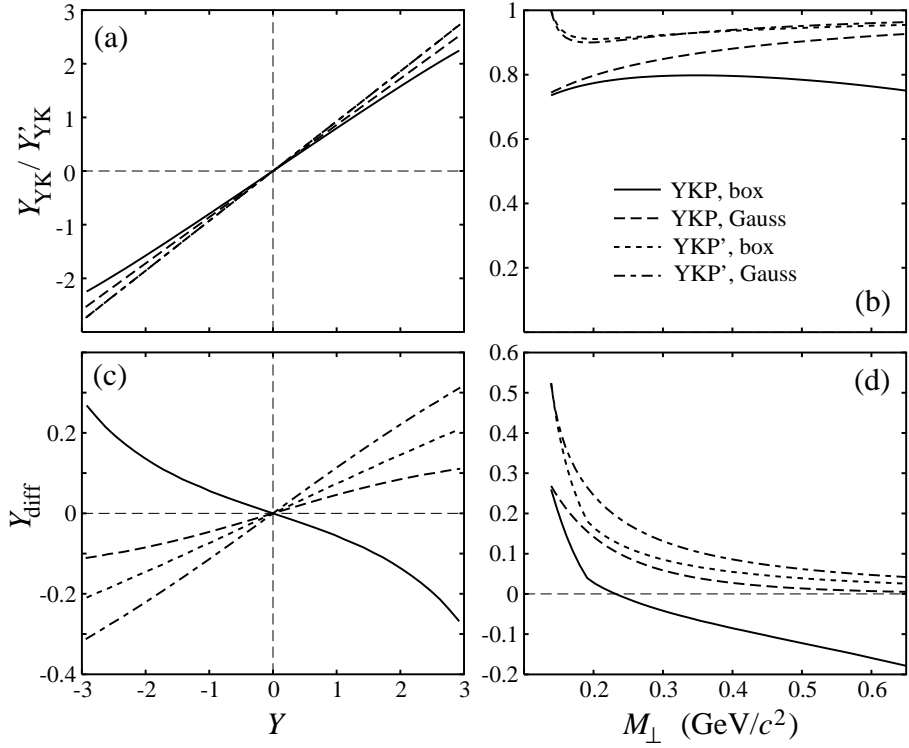}}\end{center}
\caption{{\bf Figure 4\ \ } Investigation of the YK/YK' rapidities resulting 
from the two studied models: box-shaped (solid and dotted line) and Gaussian
(dashed and dash-dotted) (see text). Quantities related to YKP parametrization
are plotted by solid and dashed lines, those related to the YKP' one by
dotted and dash-dotted lines. (a) YK/YK' rapidity as function of pair rapidity
$Y$ studied in the center-of-mass (CMS) frame of the fireball, 
$K_\perp = 0.3\, \mbox{GeV}/c$, note that dotted and dash-dotted line 
coincide;
(b) the same quantities as functions of $M_\perp$ for $Y=1$ calculated in the
source CMS ($\eta_0 = 0$); (c) the difference $Y_{\rm diff}$ between the
YK/YK' rapidity and the space-time rapidity 
of the point of maximal emissivity (maximum of the emission function) as
function of $Y$, $K_\perp = 0.3\, \mbox{GeV}/c$; (d) the same quantity but as
function of $M_\perp$ for $Y = 1$, $\eta_0 = 0$.
}
\end{figure}
The velocity parameters are again encoded via the corresponding
rapidities as introduced in (\ref{e30}). In Fig.~4a we show the 
characteristic dependence of the Yano-Koonin rapidity on the pair 
rapidity for a longitudinally (boost-invariantly) expanding source,
as observed in the source center-of-mass (CMS) frame. The modified 
YK rapidity behaves in the same way. Note that the modified
YK rapidity seems to be rather insensitive to the shape of the 
transverse density profile. This was already seen in Fig.~3c'.
Fig.~4b (which is in fact a compilation of Figs.~3c and 3c') shows 
the $M_\perp$-dependence of these rapidities for $Y = 1$ in the CMS. 

As discussed in the previous section, it is important to check to what 
extent the Yano-Koonin velocity really coincides with the velocity of 
the effective source. We also discussed the ambiguity of defining the 
latter for expanding effective sources. Here we study the difference 
$Y_{\rm diff}$ between the YK and/or Modified YK rapidity and the 
space-time rapidity (which is also the longitudinal fluid rapidity) 
of the point of maximal emissivity $\eta_{\rm max}$ (maximum of the 
emission function). While for most studied cases this difference 
is positive for $\Yyk>0$ and negative in the opposite case (i.e. the 
YK/Modified YK velocity is ``faster'' than the fluid velocity at
the maximum emissivity point), it is the other way around for high 
$M_\perp$ pions from the box-shaped source. This difference is therefore
again a model-dependent feature. However, it is important to stress 
that in all cases $Y_{\rm diff}$ is considerably smaller than $\Yyk$ 
or $\Ymyk$. Note also that, as discussed before, the rapidity 
$\eta_{\rm max}$ may well not be the ``right'' one to express the 
rapidity of effective source: if the source expands there is always an 
ambiguity of how to define its rapidity. Our choice is one of several
posibilities. The small value of $Y_{\rm diff}$ should thus not be 
taken quantitatively as the difference between the experimentally 
accessible quantity $\Yyk$ and the true source rapidity; its smallness
rather indicates that the former maps the latter very well. We conclude 
that the measurement of longitudinal expansion in this way is reliable and
does not depend on particular assumptions about the transverse geometry. 
This is valid for both YKP and YKP' parametrizations. 

\subsection{The emission duration}
\label{duration}

In order to analyze the measurement of the emission duration in more 
detail, we show in Fig.~5 the relevant correlation radii of all 
three parametrizations studied in this paper,
\begin{figure}[t]
\begin{center}\mbox{\input epsf \epsfysize 6.75cm
                        \epsfbox{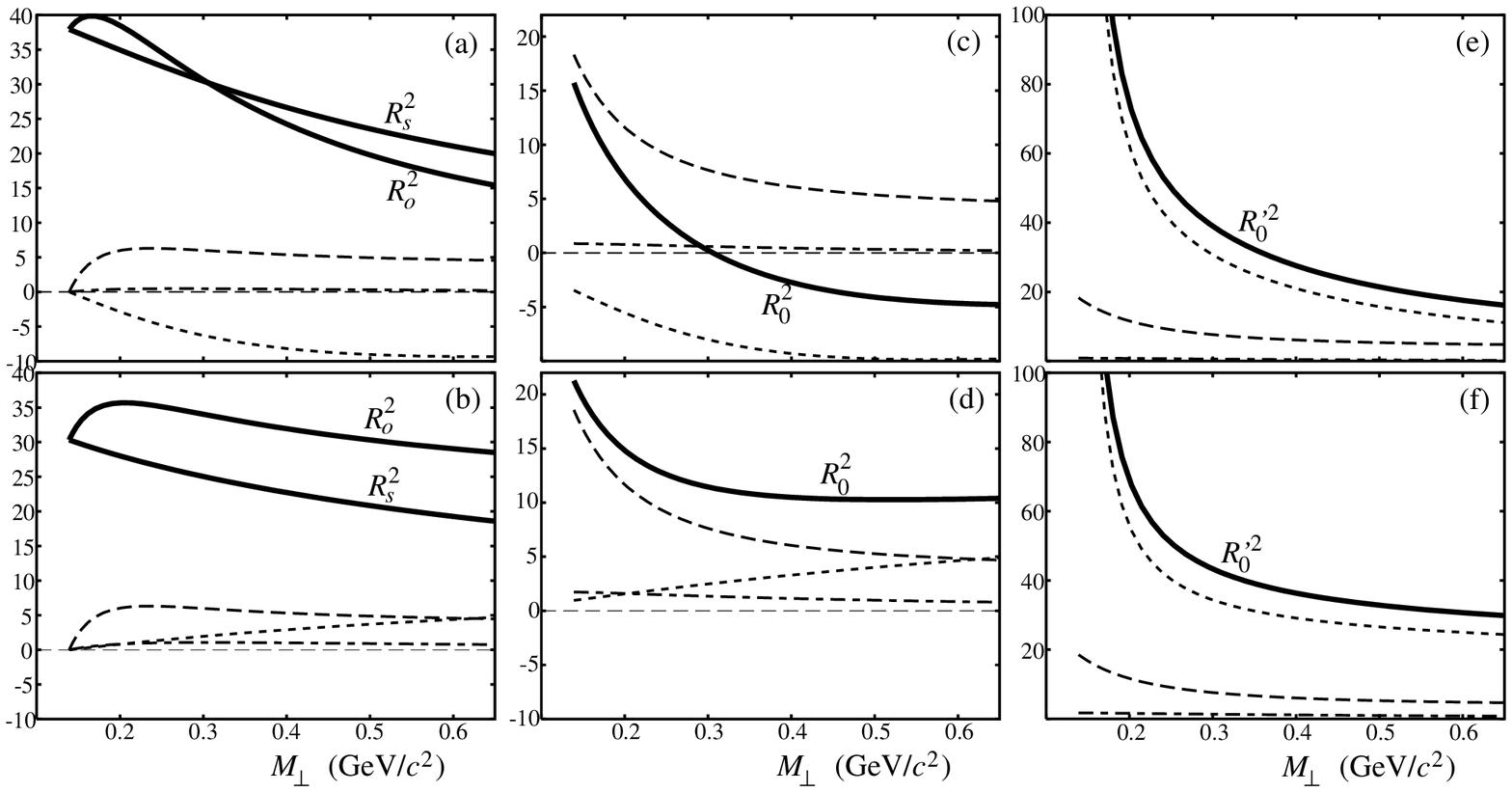}}\end{center}
\caption{{\bf Figure 5\ \ } Correlation radii related to the emission 
duration (plotted by thick solid lines) as functions of $M_\perp$. 
Upper row: box-shaped source, lower row: Gaussian source (see text). 
The remaining curves show the individual contributions: (a) and (b): 
$\bp^2 \la \tct^2\ra$ (dashed), $\la \tcx^2 - \tcy^2 \ra$ (dotted) 
and $-2\bp\la \tcx \tct \ra$ (dash-dotted); (c) and (d): same as before 
but all terms divided by $\bp^2$; (e) and (f): same as (c) and (d) 
but the dotted line corresponds to $\la \tcx^2 \ra/\bp^2$. Units 
on the ordinate are fm${}^2$.}
\end{figure}
together with the individual contributions out of which they are
constructed. This study was performed for pions at $Y = \eta_0 = 0$; in 
this case the LCMS and YK frame coincide and the decomposition is easy. 
This Figure can be summarized as follows: the geometric contribution 
$\la \tcx^2 - \tcy^2 \ra$ is at least as important as the temporal 
contribution $\la \tct^2 \ra$. This is even more true for the box-shaped 
source: here $\Rod < \Rsd$ and $\Rnd < 0$. The reason for this is the rapid
decrease of the size in outward direction which was illustrated in
Fig.~1. Note that such a behaviour was suggested in \cite{HV98} as a 
possible indicator for opacity of the source, i.e., surface-dominated
emission. We see here that already a source with a box-shaped transverse 
density profile exhibits this ``opacity signature'' for large $K_\perp$, 
as clearly seen in Fig.~1. For us, however, it is important to conclude 
that, unless $c^2\la \tct^2 \ra$ is considerably larger than the 
geometric size of the source, the lifetime cannot be reliably extracted 
from interferometric measurement as one cannot really correct for 
contributions from the transverse geometry. There exist many 
combinations of $\la \tct^2 \ra$ and $\la \tcx^2 - \tcy^2 \ra$ which 
all lead to the same interferometric signal.

We finally note that, as shown in Figs.~5e,f, the radius parameter 
$\Rmnd$ in the modified YKP parametrization does not allow for an 
extraction of the source lifetime, as already anticipated in 
Section \ref{meas}. $\Rmnd$ is seen to be completely dominated by 
the term $\la \tcx^2 \ra/\bp^2$ and practically insensitive to 
$\la \tct^2 \ra$.  

\section{Conclusions}
\label{conc}

We have shown that the recently advocated YKP parametrization is not defined
for all choices of kinematic variables. It is important to stress that this
non-definiteness can only be identified experimentally through comparison of
fit results from different parametrizations via the relations introduced in
Appendix~\ref{cross}. As it is hard to interpret fit results from a possibly
ill-defined parametrization it is absolutely necessary to perform such 
comparisons and cross-checks, and one should worry seriously about cases 
where inconsistencies appear!

A Modified YKP parametrization was introduced which does not suffer
from these problems. Eventually, it therefore also can serve for the check
of the YKP radius parameters extracted from a fit to the data \cite{App98}.
The corresponding relations were given in Appendix~\ref{cross}.

It was shown that the velocity parameters of both YKP and Modified
YKP parame\-tri\-za\-tions accurately reflect the longitudinal velocity of 
the effective source. This feature appears to be fairly model-independent.
The idea \cite{Pod83} of introducing these velocity parameters as the 
effective velocity of a non-expanding effective source thus remains valid
in the context of strongly expanding emitters.

Our consideration of the measurement of the emission duration showed that
it is practically impossible to extract this quantity in a model-independent
way. On the other hand, clear differences in the correlation patterns 
resulting from models with Gaussian and with box-shaped transverse 
profiles were observed. These could help to decide between models 
which are more or less suitable for a description of the data. In fact, 
recent studies show that a box-shaped source fits the data of the 
NA49 collaboration better than a Gaussian one \cite{Diss}. By making
such specific assumptions about the transverse source geometry (and only 
then) the emission duration can be extracted from the data fit. 

\medskip
{\bf Acknowledgments:} The work of BT was supported by DAAD. 
The work of UH was supported by DFG, BMBF and GSI. We acknowledge 
stimulating discussions with Daniel Ferenc and Urs Wiedemann and 
thank Josef Sollfrank for a careful reading of the manuscript.

Since this is a Festschrift contribution, BT wants to take the opportunity
to thank J\'an Pi\v{s}\'ut for the honour to belong to his students and 
friends. I have learned a lot from him and physics makes much more fun 
when taught by people like J\'an. UH would like to thank J\'an for the 
pleasure of many good scientific exchanges and collaborations, as well 
as for sending so many good young students to Regensburg for visits 
which have always been scientifically very fruitful. We join the other 
congratulants: all the best, J\'an! 


\appendix


\section{Definition range for the YKP/Modified YKP parameters}
\label{discrim}

The Yano-Koonin velocity $\vyk$ is only defined if the discriminant
\be
\label{def1}
D = (A + B)^2 - 4 C^2
\, ,
\ee
is positive. Here we study the conditions for that. 
From (\ref{e19}) it is clear that
\beaa{def2} \bea
A' & \geq & 0 \, ,  \label{def2a} \\
B' & \geq & 0 \, ,  \label{def2b} 
\eea \eeaa
so we conclude
\be
\label{def3}
|A' + B' | = A' + B'
\, .
\ee
It can be proven that
\be 
\label{def4}
D' = (A' + B')^2 - 4 C^{\prime 2} \geq 0 \, ,
\ee
or
\be
\label{def5}
| A' + B' | - 2 |C'| \geq 0 \, ,
\ee
i.e., the modified Yano-Koonin velocity is defined everywhere. Indeed, due to 
(\ref{def3}) inequality (\ref{def5}) may be written as
\be
\label{def6}
A' + B' \mp 2 C' \geq 0
\, ,
\ee
where the upper sign stands for $C' > 0$ and the lower for $C' < 0$.
Inserting expressions (\ref{e19}) leads to
\be
\label{def7}
A' + B' \mp 2 C' =
\lla \left \{ \left ( \tct - \frac{1}{\bp} \tcx \right ) \mp
\left (\tcz - \frac{\bl}{\bp} \tcx \right ) \right \}^2 \rra \geq 0
\, .
\ee
This proves (\ref{def4}) -- the modified YKP parametrization is defined
everywhere (except, of course, the point $K_\perp = \bp = 0$ 
\cite{YKPlett,YKPlong}).

From expressions (\ref{e19}) one also can see that
\bea
A + B & = & A' + B' - \frac{1 + \bl^2}{\bp^2} \la \tcy^2 \ra
\, ,  \label{def8}\\
2 C & = & 2 C' - \frac{2 \bl}{\bp^2} \la \tcy^2 \ra \, .
\label{def9} 
\eea
It is therefore possible to obtain negative values for $D$:
\be
\label{def10}
D = (A + B )^2 - 4 C^2 < 0
\, .
\ee
For example, if $\bl = 0$ and $ \la \tcy^2 \ra$ has a value bigger than
$ \la \tcx^2 \ra$, there exist values for $\bp$ such that
\be
\label{def11}
| A + B | = | A' + B' - \frac{1}{\bp^2} \la \tcy^2 \ra |
< 2 |C'| = 2 |C|
\, .
\ee
Hence, there are kinematic regions where the YKP parametrization is not
defined. (For example see Section~\ref{exam}.)

\section{Relations between correlation radii of different
parametrizations}
\label{cross}

In this Appendix relations between correlation radii of various
parametrizations are listed. 

The simplest and most obvious one is 
\be
\label{Rs-p-mp}
\Rsd = \Rtd = \Rmtd \, .
\ee
The remaining YKP parameters are related to the modified ones in a simple
way:
\beaa{MYKP2YKP} \bea
A & = & \gamma^{\prime 2}\left ( \Rmnd + \vmykd \Rmpd \right) -
	\frac{1}{\bp^2}\, \Rmtd\, , \label{MYKP2YKPa} \\
B & = & \gamma^{\prime 2} \left ( \Rmpd + \vmykd \Rmnd\right ) -
	\frac{\bl^2}{\bp^2} \Rmtd \, , \label{MYKP2YKPc} \\
C & = &  \gamma^{\prime 2} \vmyk \left ( \Rmnd + \Rmpd \right ) -
	\frac{\bl}{\bp^2} \Rmtd \, , \label{MYKP2YKPd}
\eea \eeaa
and the inverse relations are given by
\beaa{YKP2MYKP} \bea
A' & = & \gamma^2 \left ( \Rnd + \vykd \Rpd \right ) + \frac{1}{\bp^2}
	\Rtd \, , \label{YKP2MYKPa} \\	
B' & = & \gamma^2 \left ( \Rpd + \vykd \Rnd \right ) +
	\frac{\bl^2}{\bp^2} \Rtd \, , \label{YKP2MYKPb} \\
C' & = & \gamma^2 \vyk \left ( \Rnd + \Rpd \right ) +
	\frac{\bl}{\bp^2} \Rtd \, . \label{YKP2MYKPc}
\eea \eeaa
Similar relations exist, of course, between the original and/or
modified YKP parameters and the Cartesian correlation radii. The latter
are expressed in terms of YKP parametrization \cite{YKPlett,YKPlong}
\beaa{YKP2BP} \bea
\Rod - \Rsd \equiv R_{\rm diff}^2 & = & \bp^2 \gamma^2 \left ( \Rnd +
	\vykd \Rpd \right ) \, , \label{YKP2BPa} \\
\Rld & = & (1 - \bl^2 ) \Rpd + \gamma^2 (\bl - \vyk)^2 \left ( \Rnd
	+ \Rpd\right )\, , \label{YKP2BPb} \\
\Rold & = & \bp \left ( -\bl \Rpd + \gamma^2 (\bl-\vyk) \left ( \Rnd + 
	\Rpd \right ) \right ) \, , \label{YKP2BPc}
\eea \eeaa
while the inverse relations read
\beaa{BP2YKP} \bea
A & = & \frac{1}{\bp^2} R_{\rm diff}^2 \, , \label{BP2YKPa} \\
B & = & \Rld - 2\frac{\bl}{\bp} \Rold + \frac{\bl^2}{\bp^2}
	R_{\rm diff}^2	\, , \label{BP2YKPb} \\
C & = & -\frac{1}{\bp} \Rold + \frac{\bl}{\bp^2} R_{\rm diff}^2 \, .
	\label{BP2YKPc} 
\eea \eeaa
If instead of the original one the Modified YKP parametrization is used then
\beaa{MYKP2BP} \bea
\Rod & = & \bp^2 \gamma^{\prime 2} \left ( \Rmnd +
	\vmykd \Rmpd \right ) \, , \label{MYKP2BPa} \\
\Rld & = & (1 - \bl^2 ) \Rmpd + \gamma^{\prime 2} (\bl - \vmyk)^2 \left ( \Rmnd
	+ \Rmpd\right )\, , \label{MYKP2BPb} \\
\Rold & = & \bp \left ( -\bl \Rmpd + \gamma^{\prime 2} (\bl-\vmyk) 
	\left ( \Rmnd + \Rmpd \right ) \right ) \, , \label{MYKP2BPc}
\eea \eeaa
and
\beaa{BP2MYKP} \bea
A' & = & \frac{1}{\bp^2} \Rod \, , \label{BP2MYKPa} \\
B' & = & \Rld - 2\frac{\bl}{\bp} \Rold + \frac{\bl^2}{\bp^2}
	\Rod	\, , \label{BP2MYKPb} \\
C' & = & -\frac{1}{\bp} \Rold + \frac{\bl}{\bp^2} \Rod \, .
	\label{BP2MYKPc} 
\eea \eeaa
Note that the only formal change in (\ref{MYKP2BP})/(\ref{BP2MYKP}) 
if compared with (\ref{YKP2BP})/(\ref{BP2YKP}) is
the replacement of $R_{\rm diff}^2$ with $\Rod$ and the use of primed 
radius parameters.


\section{Off-shell corrections to the YKP/Modified YKP correlation radii}
\label{corr}

In this Appendix we list the correction terms to the Gaussian width 
parameters of the correlator which arise from releasing the approximations 
made in (\ref{CSos}), namely:
\begin{itemize}
\item[(Ad)]	setting $P_1(p_1)P_1(p_2) \to [P_1(K)]^2$ in denominator
	of that relation -- these corrections are denoted with
	$\delta_d$;
\item[(An)] 	replacing $K$ with its on-shell value in the emission function
	appearing in the numerator, i.e., $K^0 \to E_K=\sqrt{\vec K^2 + m^2}$
	-- corresponding terms will be assigned with $\delta_n$.
\end{itemize}
Here
\be
P_1(p) = \int d^4x\, S(x,p)\, ,
\label{spec}
\ee
denotes the single-particle invariant momentum distribution. In fact, only 
the correction terms to the shorthands (\ref{e15}) and (\ref{e19}) are 
derived; those for the YKP/Modified YKP radii can then be obtained 
via (\ref{e14})/(\ref{e18}).

The following relations are obtained by combining the correction terms 
known for the Cartesian parametrization \cite{CSH95b}. They read for (Ad)
\beaa{ctCartd} \bea
\delta_d \Rsd & = & \frac{1}{4K_\perp} \frac{d}{dK_\perp} \, \ln P_1(K)\, ,
\label{ctRsd} \\
\delta_d \Rod & = & \frac{1}{4} \frac{d^2}{dK_\perp^2}\, \ln P_1(K) \, ,
\label{ctRod} \\
\delta_d \Rld & = & \frac{1}{4} \frac{d^2}{dK_l^2}\, \ln P_1(K) \, ,
\label{ctRld} \\
\delta_d \Rold & = & \frac{1}{4} \frac{d^2}{dK_\perp\, dK_l}\, \ln P_1(K) \, ,
\label{ctRoldd}
\eea \eeaa
and for (An)
\beaa{ctCartn} \bea
\delta_n \Rsd & = & -\frac 12 \frac{d}{dm^2} \, \ln P_1(K) \, , 
\label{ctRsn} \\
\delta_n \Rod & = & -\frac 12 (1-\bp^2) \frac{d}{dm^2}\, \ln P_1(K) \, ,
\label{ctRon} \\
\delta_n \Rld & = & -\frac 12 (1-\bl^2) \frac{d}{dm^2}\, \ln P_1(K) \, ,
\label{ctRln} \\
\delta_n \Rold & = & \frac 12 \bp \bl \frac{d}{dm^2} \, \ln P_1(K) \, .
\label{ctRoln}
\eea \eeaa

Now we have to combine these terms according to (\ref{BP2YKP}) and 
(\ref{BP2MYKP}). The most trivial correction, of course, is the one 
belonging to $\Rtd$ and $\Rmtd$ as these radius parameters are identical 
with $\Rsd$:
\be
\label{ctRtd}
\delta_d \Rtd = \delta_d\Rmtd = \delta_d \Rsd = 
\frac {1}{4K_\perp} \, \frac{d}{dK_\perp}\, \ln P_1(K) \, .
\ee
For the remaining parameters one finds
\beaa{ctYKPd} \bea
\delta_dA & = & \frac{1}{4\bp^2}\, \left ( \frac{d^2}{dK_\perp^2} -
\frac{1}{K_\perp}\, \frac{d}{dK_\perp} \right ) \, \ln P_1(K) \, ,
\label{ctAd} \\
\delta_dB & = & \frac 14 \left [ \widehat{\cal D}^2 - \frac{\bl^2}{\bp^2}\,
\frac{1}{K_\perp}\, \frac{d}{dK_\perp} \right ] \ln P_1(K) \, ,
\label{ctBd} \\
\delta_d C & = & - \frac 14 \left [ \widehat{\cal D} + \frac{\bl}{\bp}\, 
\frac{1}{K_\perp} \right ] \, \frac{d}{dK_\perp}\, \ln P_1(K) \, ,
\label{ctCd}
\eea \eeaa
where two operators have formally been introduced 
\beaa{ag1} \bea
\widehat{\cal D} & = & \frac{d}{dK_l} - \frac{\bl}{\bp}\, \frac{d}{dK_\perp}
\, , \label{ag1a} \\
\widehat{\cal D}^2 & = & \frac{d^2}{dK_l^2} - \frac{2\bl}{\bp} \, 
\frac{d^2}{dK_l\, dK_\perp} + \frac{\bl^2}{\bp^2} \, \frac{d^2}{dK_\perp^2}
\, . \label{ag1b}
\eea \eeaa
Corresponding relations for the modified radii read
\beaa{ctMYKPd} \bea
\delta_d A' & = & \frac{1}{4\bp^2} \, \frac{d^2}{dK_\perp^2}\, \ln P_1(K)
\, , \label{ctApd} \\
\delta_d B' & = & \frac{1}{4}\,  \widehat{\cal D}^2 \, \ln P_1(K) \, ,
\label{ctBpd} \\
\delta_d C' & = & - \frac{1}{4} \, \frac{1}{\bp} \, \widehat{\cal D}\, 
\frac{d}{dK_\perp} \, \ln P_1(K) \, . \label{ctCpd}
\eea \eeaa

On the first sight, these correction terms diverge as $\bp \to 0$. In fact,
such a divergence cannot be {\em a priori} excluded in 
(\ref{e15})/(\ref{e19}). One might argue that $\la \tcx^2 - \tcy^2 \ra$, 
$\la \tct\tcx \ra$, and $\la \tcz \tcx \ra $ in (\ref{e15}) should vanish 
for $\bp \to 0$ due to restored azimuthal symmetry; it is, however, not 
guaranteed that they vanish sufficiently fast. Fortunately, the apparent
singularities actually cancel: rewriting 
$M_\perp = \sqrt{K_\perp^2 + m^2}$
\beaa{ag2} \bea
\frac{d}{dK_\perp} & = & \frac{K_\perp}{M_\perp}\, \frac{d}{dM_\perp} \, ,
\label{ag2a} \\
\frac{d^2}{dK_\perp^2} & = & \frac{m^2}{M_\perp^3}\, \frac{d}{dM_\perp}
+ \frac{K_\perp^2}{M_\perp^2}\, \frac{d^2}{dM_\perp^2} \, ,
\label{ag2b}
\eea \eeaa
we find
\beaa{ctYKPdm} \bea
\delta_d A & = & \frac{E_K^2}{4\, M_\perp^2} \, \left [ 
\frac{d^2}{dM_\perp^2} - \frac{1}{M_\perp} \, \frac{d}{dM_\perp} \right ]
\, \ln P_1(K) \, , \label{ctAdm} \\
\delta_dB & = & \frac{1}{4} \, \left [ \widehat {\cal D'}^2 -
\frac{\bl E_K^2}{M_\perp^3}\, \frac{d}{dM_\perp} \right ] \, \ln P_1(K)
\, , \label{ctBdm} \\
\delta_d C & = & - \frac{1}{4}\, \frac{E_K}{M_\perp} \, \left [
\widehat {\cal D'} + \frac{\bl E_K}{M_\perp^2} \right ] \, 
\frac{d}{dM_\perp} \, \ln P_1(K) \, , \label{ctCdm} 
\eea \eeaa
and
\beaa{ctMYKPdm} \bea
\delta_dA' & = & \frac{E_K^2}{4\, M_\perp^2} \, \left [
\frac{d^2}{dM_\perp^2} + \frac{1}{\bp^2}\, \frac{m^2}{E_K^2\, M_\perp}
\frac{d}{dM_\perp} \right ] \, \ln P_1(K) \, , \label{ctApdm} \\
\delta_dB' & = & \frac {1}{4} \, \left [ \widehat {\cal D'}^2 +
\frac{\bl^2}{\bp^2}\, \frac{m^2}{M_\perp^3}\, \frac{d}{dM_\perp} \right ]
\, \ln P_1(K) \, , \label{ctBpdm} \\
\delta_d C' & = & - \frac{1}{4}\, \frac{E_K}{M_\perp} \, \left [
\widehat {\cal D'} - \frac{\bl}{\bp^2}\, \frac{m^2}{E_K\, M_\perp^2}\,
\frac{d}{dM_\perp} \right ] \, \ln P_1(K) \, , \label{ctCpdm}
\eea \eeaa
with
\beaa{ag3} \bea
\widehat {\cal D'} & = & \frac{d}{dK_l} - \frac{\bl\, E_K}{M_\perp}\,
\frac{d}{dM_\perp} \, , \label{ag3a} \\
\widehat {\cal D'}^2 & = & \frac{d^2}{dK_l^2} - \frac{2\, \bl\, E_K}{M_\perp}
\, \frac{d^2}{dK_l\, dM_\perp} + \frac{\bl^2 \, E_K^2}{M_\perp^2} \,
\frac{d^2}{dM_\perp^2} \, . \label{ag3b}
\eea \eeaa
From (\ref{ctYKPdm}) one sees that the correction terms belonging to
the YKP parametrization behave well. The singularities cancel thanks to
the appearance of $R_{\rm diff}^2$ in (\ref{BP2YKP}). In case of the 
modified parametrization $\Rod$ appears instead of $R_{\rm diff}^2$  
in (\ref{BP2MYKP}), and thus the correction terms (\ref{ctMYKPdm}) still
diverge for vanishing $\bp$.

Corrections due to approximation (An) are obtained in a similar way as
above. The one corresponding to $\Rtd$ (and $\Rmtd$ as well) is again obvious 
\be
\label{ctRt2n}
\delta_n \Rtd = \delta_n \Rmtd = \delta_n \Rsd = -\frac{1}{2} \,
\frac{d}{dm^2} \, \ln P_1(K) \, ,
\ee
and the relations for YKP parametrization are
\beaa{ctYKPn} \bea
\delta_n A & = & \frac{1}{2} \, \frac{d}{dm^2} \, \ln P_1(K) \, ,
\label{ctAn} \\
\delta_n B & = & - \frac{1}{2} \, \frac{d}{dm^2} \, \ln P_1(K) \, ,
\label{ctBn} \\
\delta_n C & = & 0 \, . \label{ctCn} 
\eea \eeaa
For the modified YKP parameters one finds
\beaa{ctMYKPn} \bea
\delta_n A' & = & \frac{1}{2} \, \frac{\bp^2-1}{\bp^2}\, 
\frac{d}{dm^2} \, \ln P_1(K) \, , \label{ctApn} \\
\delta_n B' & = & - \frac{1}{2} \, \left ( 1+\frac{\bl^2}{\bp^2} \right )
\frac{d}{dm^2} \, \ln P_1(K) \, , \label{ctBpn} \\
\delta_n C' & = & -\frac{1}{2}\, \frac{\bl}{\bp^2}\, \frac{d}{dm^2}\,
\, \ln P_1(K) \, . \label{ctCpn}
\eea \eeaa 
While the corrections to the YKP parameters are clearly regular, this is 
not true for the modified ones. These may be rewritten by using
\be
\label{ag4}
\frac{d}{dm^2} = \frac{1}{2}\, \frac{1}{M_\perp}\, \frac{d}{dM_\perp}\, .
\ee
This leads to 
\beaa{ctMYKPnm} \bea
\delta_n A' & = & \frac{1}{4} \, \frac{1}{M_\perp}\, \frac{\bp^2-1}{\bp^2}\,
\frac{d}{dM_\perp} \, \ln P_1(K) \, , \label{ctApnm} \\
\delta_n B' & = & -\frac{1}{4} \, \left ( 1+\frac{\bl^2}{\bp^2} \right )
\, \frac{1}{M_\perp}\, \frac{d}{dM_\perp}\, \ln P_1(K) \, , 
\label{ctBpnm} \\
\delta_n C' & = & - \frac{1}{4} \, \frac{\bl}{\bp^2} \, \frac{1}{M_\perp}\,
\frac{d}{dM_\perp} \,\ln P_1(K) \, . \label{ctCpnm}
\eea \eeaa
If now these corrections are added to those due to (Ad) displayed in 
(\ref{ctMYKPdm}), one finally arrives at the total correction terms to the
modified YKP radii:
\beaa{ctMYKP} \bea
\delta A' & = & \frac{1}{4} \, \frac{1}{M_\perp} \, \left [
	\frac{E_K^2}{M_\perp}\, \frac{d^2}{dM_\perp^2} + 
	\left ( 1 - \frac{E_K^2}{M_\perp^2} \right ) \frac{d}{dM_\perp} 
	\right ] \, \ln P_1(K) \, , \label{ctAp} \\
\delta B' & = & \frac{1}{4} \left [ \widehat{\cal D'}^2 - \frac{1}{M_\perp}\,
	\left ( 1 + \frac{\bl^2\, E_K^2}{M_\perp^2}\right )\, 
	\frac{d}{dM_\perp} \right ] \, \ln P_1(K) \, , \label{ctBp} \\
\delta C' & = & - \frac{1}{4} \, \frac{E_K}{M_\perp} \left [
	\widehat{\cal D'} + \frac{\bl\, E_K}{M_\perp^2} \right ] \, 
	\frac{d}{dM_\perp} \, \ln P_1(K) \, . \label{ctCp}
\eea \eeaa
Note that no divergences survive in the final expressions.


\end{document}